\documentclass[prl,twocolumn,amsmath,amssymb,aps,floatfix]{revtex4-1}

\usepackage{graphicx}
\usepackage{dcolumn}
\usepackage{bm}
\usepackage{hyperref}
\usepackage[usenames,dvipsnames]{xcolor}
\usepackage[normalem]{ulem} 
\usepackage{amssymb}

\begin{document}

\title{Abrupt transition between three and two-dimensional quantum turbulence}

\author{Nicol\'as P.~M\"uller$^{1,3}$, Marc-Etienne Brachet$^2$, Alexandros 
    Alexakis$^2$,  and Pablo D.~Mininni$^1$}
\affiliation{$^1$Universidad de Buenos Aires, Facultad de Ciencias Exactas y
    Naturales, Departamento de F\'\i sica, \& IFIBA, CONICET, Ciudad
    Universitaria, Buenos Aires 1428, Argentina.}
\affiliation{$^2$Laboratoire de Physique de l'\'Ecole Normale Sup\'erieure, 
    ENS, Universit\'e PSL, CNRS, Sorbonne Universit\'e, Universit\'e de Paris, 
    F-75005 Paris, France}
\affiliation{$^3$Universit\'e C\^ote d’Azur, Observatoire de la C\^ote d’Azur, CNRS, Laboratoire Lagrange, Nice, France}

\date{\today}

\begin{abstract}
We present numerical evidence of a critical-like transition in an out-of-equilibrium mean-field description of a quantum system. By numerically solving the Gross-Pitaevskii equation we show that quantum turbulence displays an abrupt change between three-dimensional (3D) and two-dimensional (2D) behavior. The transition is observed both in quasi-2D flows in cubic domains (controlled by the amplitude of a 3D perturbation to the flow), as well as in flows in thin domains (controlled by the domain aspect ratio) in a configuration that mimics systems realized in laboratory experiments. In one regime the system displays a transfer of the energy towards smaller scales, while in the other the system displays a transfer of the energy towards larger scales and a coherent self-organization of the quantized vortices.
\end{abstract}

\maketitle

The phenomena of condensation and phase transitions in statistical mechanics has traditionally been associated with equilibria. However, observations of turbulence in experiments of gaseous Bose-Einstein condensates (BECs) \cite{henn2009emergence, white2014vortices, navon2016emergence} and of superfluid $^4$He \cite{Vinen02, Skrbek12, Fonda14} have shown that these out-of-equilibrium systems can also display multiple phases.
In particular, recent BEC experiments close to a two-dimensional (2D) regime, instead of a tendency towards disorder, display an intriguing out-of-equilibrium self-organization and the nucleation of quantized vortices \cite{Seo17, Gauthier1264, Johnstone1267} (see \cite{simula2014emergence, billam2014onsager} for numerical studies).

In classical turbulence, a reminiscent process can take place when flows are 2D. Under certain conditions, the kinetic energy can undergo an inverse cascade (moving to larger scales), and eventually create a condensate \cite{rhk_montgo}. This condensation is of a different nature than a BEC as it involves the kinetic energy of the system instead of its mass density. 
In classical three-dimensional (3D) turbulence, recent developments indicate that this far-from-equilibrium system can change its behavior as its dimensionality is changed \cite{celani2010turbulence, benavides2017critical, Alexakis2018CascadesFlows, van2019condensates} (or, equivalently, as one of its spatial dimensions is compactified, see \cite{celani2010turbulence}, and \cite{gregory1993} for an example of a transition under compactification in gravitational theories). In classical fluids, when the flow is 3D energy undergoes a direct cascade (moving to smaller scales), while as the domain that contains the fluid is made thiner, the system becomes 2D and displays an inverse cascade after a critical second-order transition.

Both classical and quantum turbulence involve non-linear and complex spatio-temporal dynamics of fields, and cascade-like solutions can develop in many different cases. In this letter we address the following questions: Is there a transition in the behavior of quantum turbulence as its dimensionality is changed as reported in recent quantum turbulence experiments \cite{Seo17, Gauthier1264, Johnstone1267}? And is this transition associated with the emergence of different out-of-equilibrium self-similar regimes (i.e., associated with a change in the direction of the energy cascade)? To this end, we study numerically 3D condensates in periodic boundary conditions using the Gross-Pitaevskii equation (GPE), exploring two configurations. In one, we solve the equations in a cubic domain and perturb an initial 2D random array of quantized vortices with a 3D perturbation, varying the amplitude of the perturbation as a control parameter. In the other, we consider a quasi-2D array of quantized vortices and vary the aspect ratio of the domain, compactifying one of its dimensions. In both cases we find evidence of an abrupt transition towards a regime that displays two-dimensionalization, spatial aggregation of quantized vortices, and inverse energy flux.

\begin{figure*}
  \includegraphics[width=.49\linewidth]{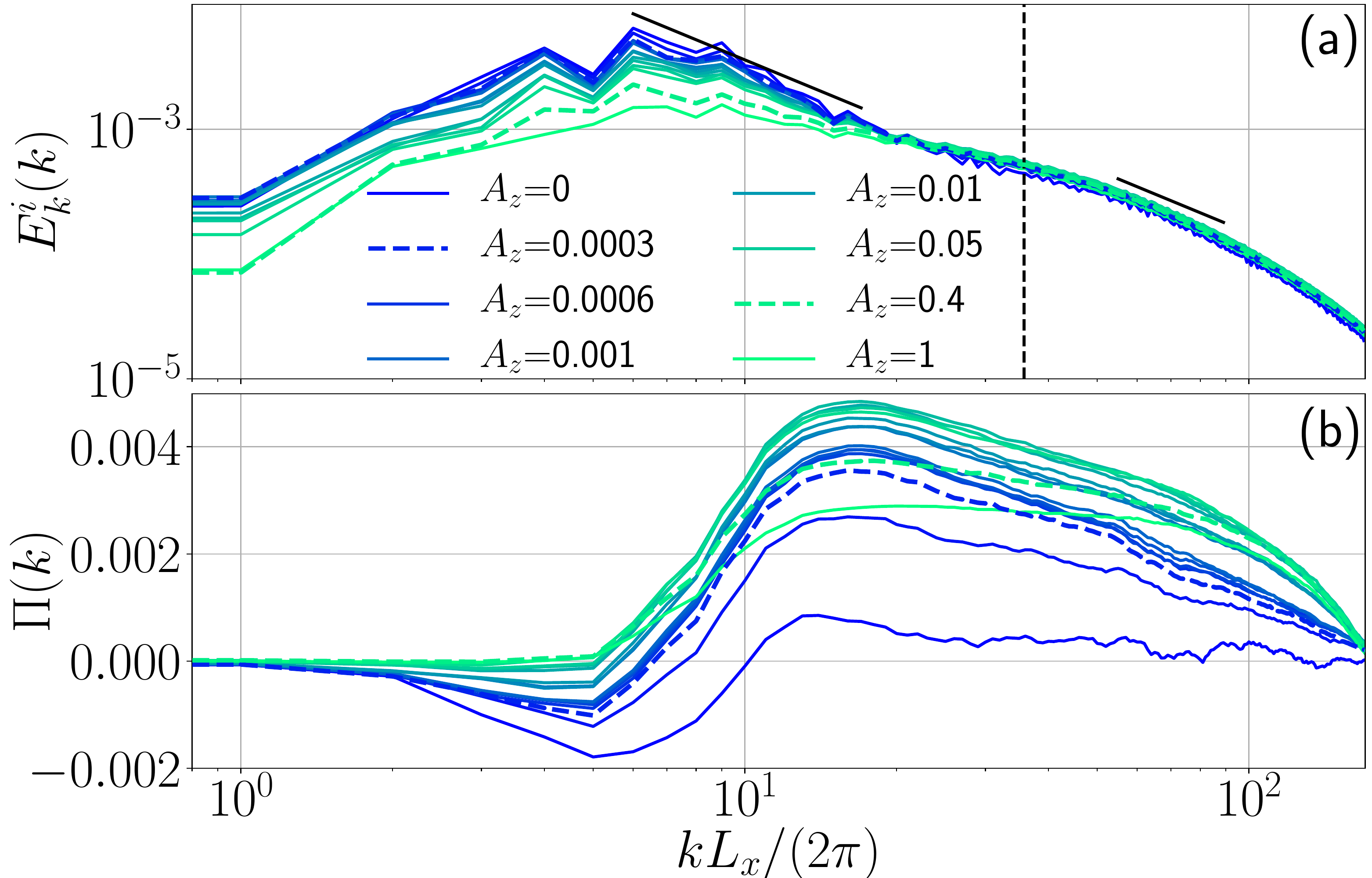}
  \includegraphics[width=.49\linewidth]{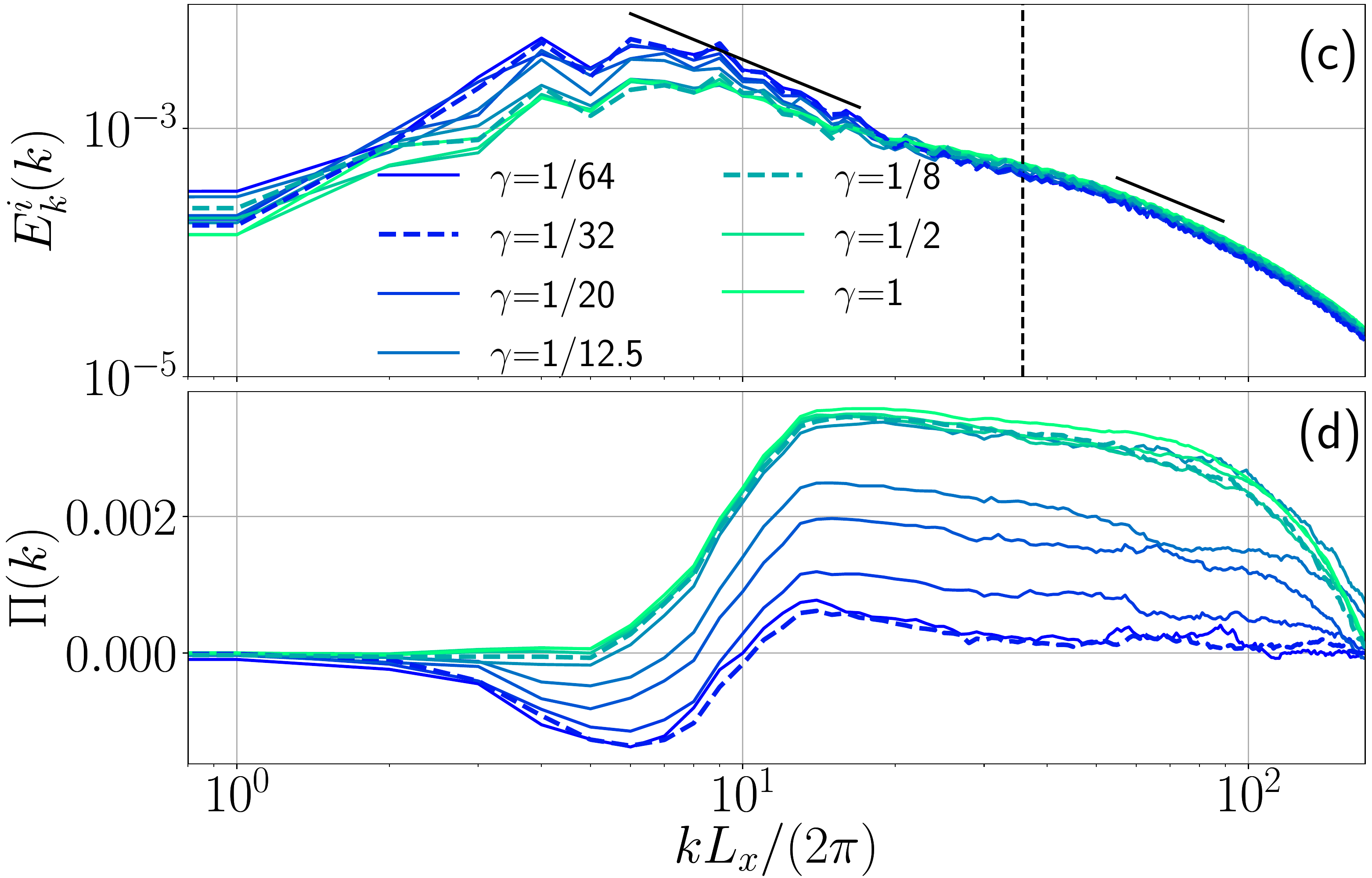}
  \caption{(a) Spectrum of the incompressible kinetic energy averaged between $t = 1$ and $10$ for simulations in cubic domains ($N_x=N_y=N_z=512$) and different values of $A_z$. Kolmogorov power laws $\sim k^{-5/3}$ are indicated as a reference by solid black lines. The vertical dashed line indicates the inverse mean intervortex distance. Note the growth of energy and a $\sim k^{-5/3}$ scaling for $k\lesssim 10$ when $A_z$ is small. (b) Total energy fluxes for the same simulations. For small $A_z$ the flux becomes negative for $k\lesssim 10$, and the positive flux for $k>10$ decreases. References for the values of $ A_z $ in (a) and (b) are provided in the inset. (c) Same as in (a) for simulations in thin domains, for different values of $\gamma$. (d) Same as in (c) for simulations in thin domains; the inset gives the values of $\gamma$. In all panels, dashed curves highlight simulations for which movies are available in \cite{sm}.}
    \label{fig:flux_cubic}
\end{figure*}

To describe the dynamics of weakly interacting bosons
of mass $m$ 
at zero-temperature we solve numerically the GPE \cite{sm},
$i \hbar \partial_t \psi = -\hbar^2 \nabla^2 \psi/(2m) + g |\psi|^2 \psi,$ where $\psi$ is the condensate wave function and $g$ is proportional to the scattering length. The fluid density, velocity, and quantized vortices can be obtained from $\psi$ using Madelung's transformation (see \cite{sm}). The GPE is solved using a parallel pseudospectral method \cite{ghost, ClarkdiLeoni2017DualTurbulence}. To achieve the largest possible scale separation (at a fixed spatial resolution), we resort to periodic boundary conditions in a 3D domain of size $L_x \times L_y \times L_z$, with spatial resolution $N_x \times N_y \times N_z$. The size of the domain is $L_x = L_y = L_\perp = 2\pi$ in dimensionless units in all cases, and $L_z = \gamma L_\perp$ where $\gamma$ is the domain aspect ratio. In these domains, we prepared a set of randomly distributed 2D vortices with a small 3D perturbation of amplitude $A_z$, such that the wavefunction is a solution of the GPE, and that the incompressible kinetic energy of the system peaks at an intermediate wavenumber $k_0\approx 10$ (i.e., the correlation length of the flow is $\ell_0 \approx L_\perp/10$; see \cite{sm} for more details on the preparation of the initial conditions and for the definition of the incompressible kinetic energy). This results on quantized vortices with a random separation, and that are perfectly 2D for $A_z=0$ while they display stronger curvature in $z$ for increasing $A_z$.

As previously mentioned, we consider two ways to observe a transition between 2D and 3D flows using these initial conditions. One of them consists on varying the amplitude of the 3D perturbation $A_z$ between $0$ and $1$ in a cubic domain. The other, is to vary the aspect ratio of the domain for fixed $A_z$ (the 2D limit case being that in which $ \gamma = L_z / L_\perp \rightarrow 0$, and the 3D case when $ \gamma = 1 $). In each case, when varying the control parameters between their two limits, classically we can expect an inverse cascade of energy in the 2D regime, and the absence thereof in the 3D case. To identify the direction of the cascades we consider two quantities: the incompressible kinetic energy spectrum $E_k^i(k)$ (see \cite{sm, Nore1997,PhysRevA.99.043605} for a detailed description of energy components in the GPE) and the total energy flux $\Pi(k) = -dE^<(k)/dt$, where $E^<(k)$ is the total energy of the system integrated up to wavenumber $k$, $E^<(k) = \int_0^k E(k') dk'$, and where $E(k)$ is the total energy spectrum \cite{sm}. A direct cascade of energy corresponds to the development of a power law in $E_k^i(k)$ for $k>k_0$ and with $\Pi(k)>0$ constant in a range of wavenumbers, while an inverse cascade of energy corresponds to a growth of $E_k^i(k)$ for $k<k_0$ and with $\Pi(k)<0$. As the system has no external steering force (but no dissipation either), an inverse cascade can only develop for a transient time \cite{Mininni2013InverseTurbulence}, and in the following we will focus on time averages of these quantities between $t = 1$ and $10$ flow turnover times, as well as on their time evolution over the same time span (with the turnover time defined as $\tau=\ell_0/U$, with $U$ the r.m.s.~initial flow velocity).

In cubic domains ($\gamma=1$) we performed two sets of simulations, with spatial resolutions of $N_x \times N_y \times N_z=256^3$ and $512^3$ grid points, varying the amplitude of the 3D perturbation $A_z$. For large values of $A_z$ the flow quickly evolves into a 3D regime, with quantized vortices rapidly being deformed, while for small $A_z$ there is a long transient in which the flow remains quasi-2D (see the videos in \cite{sm}). Figure \ref{fig:flux_cubic}(a) shows the time average of $E_k^i(k)$ for the simulations with $512^3$ grid points, and for different values of $A_z$. For large values of $A_z$ initial vertical gradients in the quantized vortices are large, and the spectrum peaks at $k\approx k_0$ followed by a spectrum compatible with a direct energy cascade and with the emission of Kelvin-waves at wavenumbers smaller than the inverse mean intervortex distance \citep{ClarkdiLeoni2017DualTurbulence}. The energy fluxes in Fig.~\ref{fig:flux_cubic}(b), specially for $A_z=1$, are positive for all $k$ and remain approximately constant for a range of wavenumbers $k>k_0$. But for small values of $A_z$ initial vertical gradients are small, and the energy spectrum grows for $k\lesssim k_0$, developing a power law compatible with Kolmogorov scaling, and with negative total energy flux for $k\lesssim k_0$ (albeit the negative flux does not remain constant with $k$, as a result of limited spatial resolution and of the inverse cascade being only transient in the absence of external forcing). In spite of this, the system develops a strong inverse transfer of energy, at least up to $t=10$. For longer times the flow eventually becomes unstable and 3D. However, we verified that the 2D behavior is not simply due to an absence of 3D motions for $A_z\ll 1$. For $t\lesssim 10$, when the systems display an inverse transfer of energy, the energy in the 3D modes for all $A_z\ne0$ is significant enough to nonlinearly act back to the 2D part of the flow and saturate its initial exponential growth, but not strong enough yet to suppress the inverse transfer.

Figures \ref{fig:flux_cubic}(c) and (d) show similar results for simulations with a fixed value of $A_z=0.1$ (such that a 3D flow with a direct energy cascade is generated when $\gamma=1$), but with different aspect ratios $\gamma$, using a spatial resolution $N_x=N_y=512$, and with $N_z$ varied between 512 and 32 grid points to keep the vertical resolution $ \Delta z$ fixed or over-resolved as $L_z$ is decreased, so that vertical gradients in the flow are always correctly resolved. Although the amplitude of the perturbation $A_z$ is fixed, by decreasing $\gamma$ we also increase the wavenumber of the vertical perturbation (i.e., vertical variations of quantized vortices increase as the domain becomes thiner). As in the cubic domain, we observe an increase in $E_k^i(k)$ for $k\lesssim k_0$ and a range of wavenumbers with $\Pi(k)<0$ but now for small values of $\gamma$, and a direct cascade of energy for large values of $\gamma$. But, unlike the case of the cubic domain, when $\gamma$ is sufficiently small the flow remains quasi-2D for very long times, and quantized vortices aggregate in physical space creating larger structures (see movies in \cite{sm}).

\begin{figure}
    \includegraphics[width=1\linewidth]{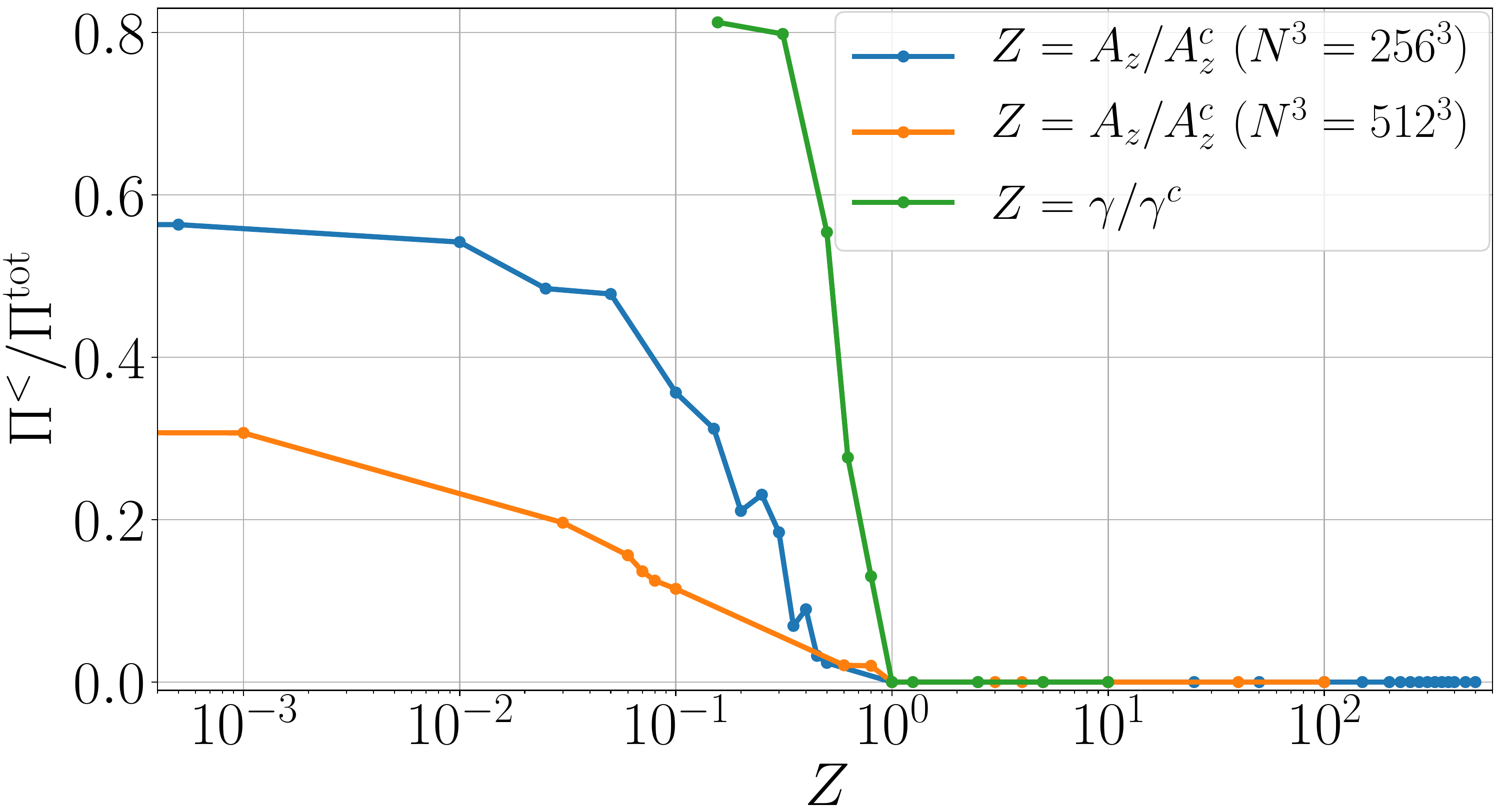}
    \caption{Ratio of inverse energy flux to total energy flux as a function of the normalized control parameter $Z$ (either $A_z$ or $\gamma$, normalized by their respective critical values $A_z^c$ or $\gamma^c$, see inset). For cubic domains two curves are shown, corresponding to spatial resolutions of $256^3$ and $512^3$ grid points.}
    \label{fig:critic_cubic}
\end{figure}

To quantify the transition between the direct and inverse cascade regimes, we need to use (as an order parameter) an observable that measures the relative strength of the inverse energy cascade. To this end we first introduce the mean inverse flux at small wavenumbers (which is zero when the flux is positive), and the mean direct flux at large wavenumbers, respectively defined as
\begin{align}
    \Pi^< &= \left| \min\left\{0,\frac{1}{k_0}\sum_{k=0}^{k_{0}} \Pi(k)\right\}\right|, \\
    \Pi^> &= \frac{1}{k_\textrm{max}-(k_0+1)}\sum_{k=k_{0}+1}^{k_\textrm{max}} \Pi(k),
    \label{eq:def_pimin}
\end{align}
where $k_\textrm{max}=N_x/3$ is the maximum resolved wavenumber in the simulations, and $ k_0 $ is as before the wavenumber where the incompressible kinetic energy is initially concentrated. We can then compute the total energy flux (in both directions) $\Pi^\textrm{tot} = \Pi^< + \Pi^> $, and define the normalized ratio of inverse energy flux to total energy flux $\Pi^</\Pi^\textrm{tot}$. Figure \ref{fig:critic_cubic} shows the behavior of this quantity for all cases studied, as a function of the amplitude of the 3D perturbation normalized by its critical value $A_z/A_z^c$ (for spatial resolutions of $256^3$ and $512^3$ grid points), and as a function of the aspect ratio normalized by its critical value $\gamma/\gamma^c$ (for fixed $A_z$). In all cases we see an abrupt change as the control parameter is varied. For $A_z/A_z^c$ or $\gamma/\gamma^c>1$ there is no inverse energy flux, while for $A_z/A_z^c$ or $\gamma/\gamma^c<1$ it grows rapidly (albeit differently in each case). In the thin domains, from Fig.~\ref{fig:flux_cubic} it can be seen that $\gamma^c \approx 0.1$, corresponding to a domain with $L_z = L_{\perp}/10 \approx 11 \xi$ (where $\xi$ is the healing length of the condensate, proportional to the vortex core radius). This implies that the 2D behavior occurs when the height of the domain is of the same order as the correlation of the initial conditions, $\gamma_c \simeq \ell_0/L_\perp$, a similar condition for the layer height and the forcing lengthscale found for the compactified case in classical flows \cite{celani2010turbulence}. In other words, a transition towards 2D behavior does not require 2D domains or very slim films. Even moderate aspect ratios are enough to trigger an inverse energy cascade.

\begin{figure}
    \centering
    \includegraphics[width=1\linewidth]{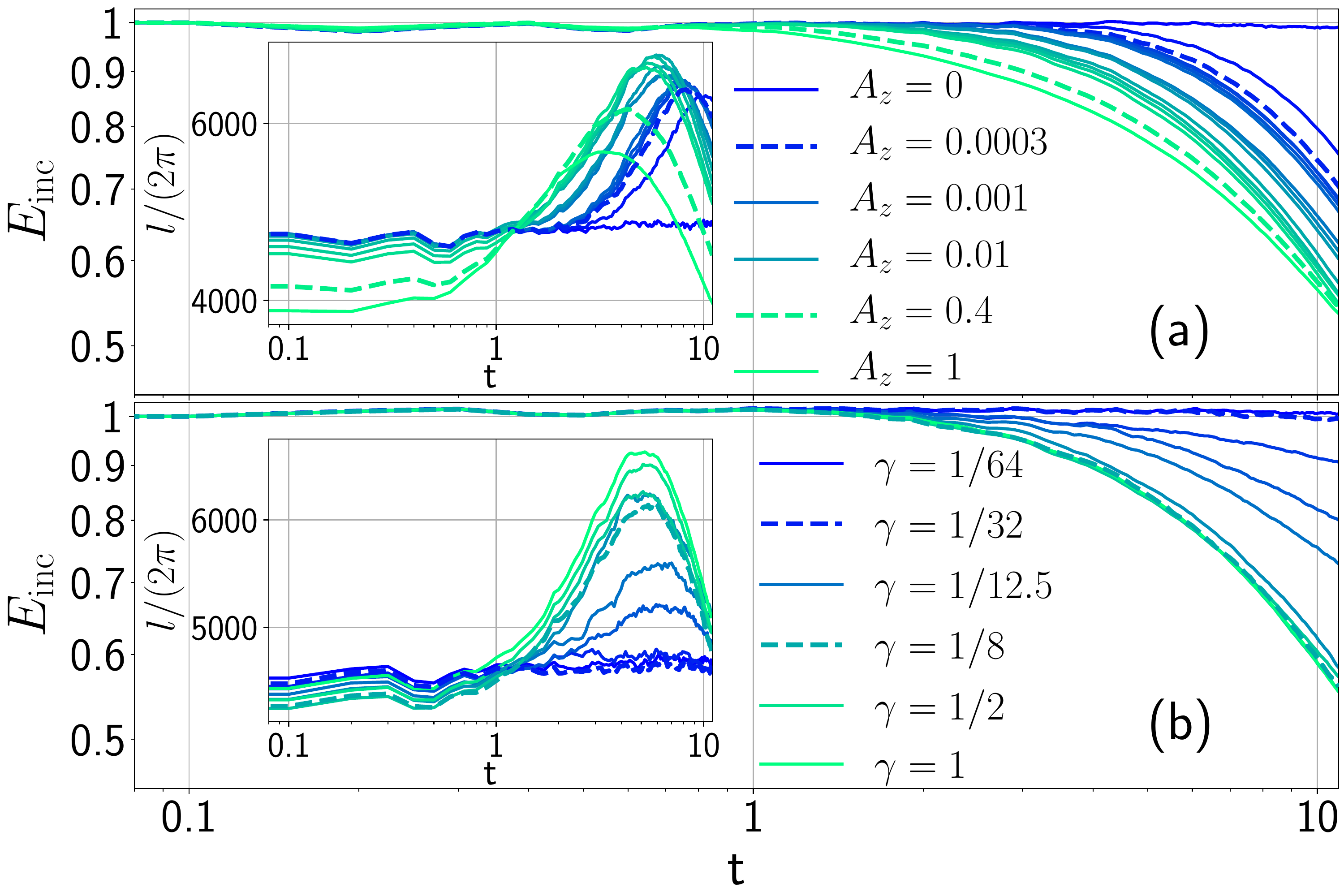}
    \label{fig:energy_evolution_cubic}
    \caption{Time evolution of the incompressible kinetic energy (a) in simulations in cubic domains with different perturbation amplitudes $A_z$ ($512^3$ runs), and (b) in simulations in domains with different aspect ratios $\gamma$. The insets show the total vortex length \cite{sm}  as a function of time for each case. References are as in Fig.~\ref{fig:flux_cubic}; a few labels are provided as guidelines.}
    \label{fig:Einc_thin}
\end{figure}

\begin{figure}
    \centering
    \includegraphics[width=1\linewidth]{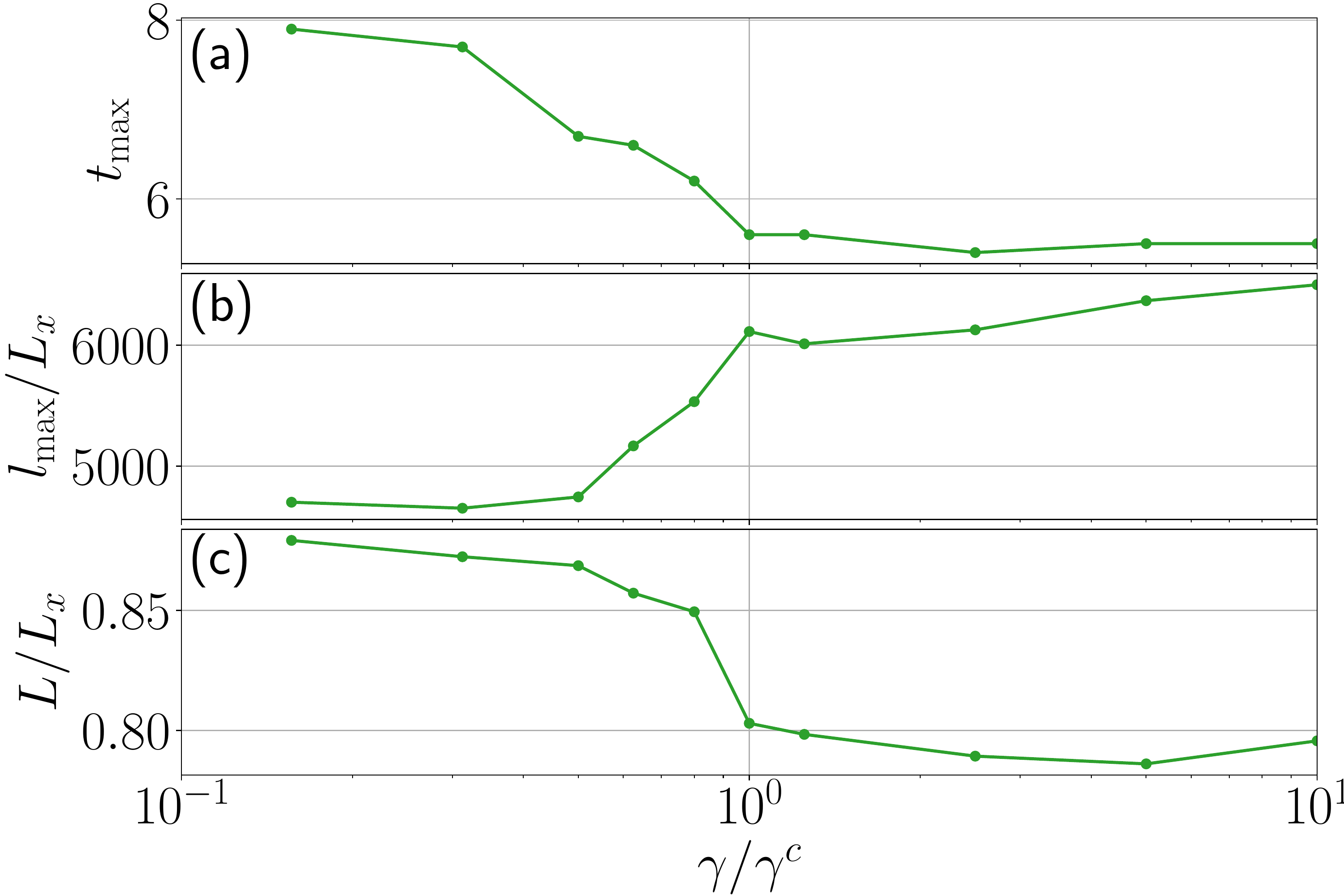}
    \caption{(a) Time to reach the maximum length of the vortices as a function of $\gamma/\gamma^c$. (b) Normalized maximum total length of the vortices, $l_\textrm{max}/L_x$, as a function of the control parameter. (c) Flow integral length scale normalized by the domain size, $L/L_x$, as a function of the same control parameter.}
    \label{fig:tmax_cubic}
\end{figure}

Energy fluxes, although they give direct indication of the presence of an inverse cascade, are not easily measurable in laboratory experiments. There are however other global quantities that are tractable experimentally and can also give an indication of a transition from 3D to 2D behavior in the flows as the control parameters are varied. Figure \ref{fig:Einc_thin} shows the incompressible kinetic energy $E_\textrm{inc}$ in these flows as a function of time, both for $512^3$ simulations in cubic boxes with different $A_z$ as well as for simulations in domains with different $\gamma$, and for each case, also the total length of the vortices as a function of time. In the simulations with large $A_z$ or $\gamma$, $E_\textrm{inc}$ decays in time after $t\approx 1$, as the direct cascade of energy transfers the incompressible kinetic energy to smaller scales where it dissipates into phonons \cite{Nore1997b,Nore1997, ClarkdiLeoni2017DualTurbulence}. However, for small $A_z$ or $\gamma$, $E_\textrm{inc}$ remains constant in time or decays very slowly, indicating energy remains at large scales as in classical 2D turbulence. The same behavior is seen in the total vortex length \cite{sm,Nore1997}, which grows and reaches a maximum in the 3D regime as a result of vortex stretching (later decaying as a result of vortex reconnection), but which remains approximately constant for all times in the cases of small $A_z$ or $\gamma$, pointing to the absence of vortex stretching as expected in 2D flows. Vortex reconnection also plays an important role at early times for large $A_z$ or $\gamma$, to speed-up the three-dimensionalization of the flow, after which vortex stretching can become more efficient. Finally, it is also important to note that in the simulations in cubic domains the total length of the vortices remains approximately constant at early times in all cases, and that the time when vortex stretching starts increases as $A_z$ decreases. This is consistent with our previous observations: In the cubic box, for smaller values of $A_z$ the flow remains quasi-2D for longer times, and the observed transient inverse cascade delays the growth of 3D excitations in the flow.

Given the above we can consider various quantities that can indicate the presence of a sharp transition. Here we focus only on the thin layer case that is somehow closer to what is experimentally realizable. Figure \ref{fig:tmax_cubic} shows the time to reach the maximum vortex stretching $t_\textrm{max}$, the maximum length of the vortices $l_\textrm{max}$, and the flow integral scale $L$ as a function of $\gamma$, which is obtained from the incompressible kinetic energy spectrum as $L = 2\pi\int k^{-1} E_k^i(k) dk / \int E_k^i(k) dk$, and provides an estimation of the flow energy containing scale. When $L\approx L_x$ (the domain size), the flow has self-organized at the largest available scale in the domain. These quantities display an abrupt change near the critical value $\gamma^c$ as $\gamma$ is varied. The time $t_\textrm{max}$ is larger when $\gamma<\gamma^c$, while the maximum vortex length is larger when $\gamma>\gamma^c$. Both behaviors are to be expected when the flow is 3D and displays vortex stretching, or when the flow is 2D and as a result does not. Finally, the flow integral scale $L$ becomes larger (and close to $L_x$) when $\gamma<\gamma^c$. This indicates that the inverse transfer of energy leads to the concentration of kinetic energy at large scales, and implies the formation of large structures in the flow (e.g., resulting from spatial aggregation of vortices). In the simulations varying $A_z$, we also verified that the overall shape of the quantities in the curves in Fig.~\ref{fig:tmax_cubic} remain the same when changing the spatial resolution of the simulations, although the actual values (e.g., the time $t_\textrm{max}$ or the maximum vortex length $l_\textrm{max}$) depend on the resolution: at larger resolution the flow becomes more turbulent and vortex stretching is more efficient.

The numerical results show the existence of a transition between 2D and 3D behavior in quantum turbulence. This transition can be obtained by varying the dimensionality of the flow (in a 3D cubic domain), or by changing the aspect ratio of the domain and compactifying one spatial dimension. Below critical values of the controlling parameters the flows display an inverse transfer of energy which results in the growth of the incompressible kinetic energy at large-scales, and in the aggregation of quantized vortices. For the quasi-2D regimes the system suffers an interesting double condensation: the BEC, and the out-of-equilibrium inverse energy cascade which can result in a condensation of the kinetic energy at the largest available scales in the system \cite{rhk_montgo}. This behavior is compatible with critical transitions reported in classical turbulence \cite{celani2010turbulence, benavides2017critical, Alexakis2018CascadesFlows, van2019condensates}, and reminiscent of recent observations in experiments of gaseous BECs \cite{Seo17, Gauthier1264, Johnstone1267}. For the 3D cubic domain, the critical amplitude of the 3D perturbation is $A_z^c \approx 10^{-2}$ (for the $512^3$ simulations), while in the thin domains the critical aspect ratio is $\gamma^c \approx 1/10$. As our system is not forced, the inverse energy cascade can only develop as a transient (see, e.g., \citep{Mininni2013InverseTurbulence} for a discussion of the equivalent configuration in the classical case), a configuration which is comparable to experiments of gaseous BECs where the flow is let to freely decay after initially stirring it \cite{henn2009emergence, white2014vortices, navon2016emergence}. However, in experiments of gaseous BECs the condensate is trapped inside a potential, which we are not considering in our simulations to increase the scale separation between the domain size and the vortex radius as much as possible. The study of the effect of trapping potentials in these cascades is left for future work.

\begin{acknowledgments}
{\it NPM and PDM acknowledge financial support from grants UBACYT No.~20020170100508BA and PICT No.~2015-3530.}
\end{acknowledgments}

\bibliography{ms}

\clearpage

\onecolumngrid
\section*{Supplemental Material: Abrupt transition between three and two-dimensional quantum turbulence}

\section{The Gross-Pitaevskii Equation} \label{sec:GPE}

In this work we study a system of weakly interacting bosons of mass $m$ at zero-temperature that is described by the GPE
\begin{equation}
    i\hbar \frac{\partial \psi}{\partial t} = -\frac{\hbar^2}{2m}\nabla^2\psi + g |\psi|^2\psi,
    \label{eq:GPE}
\end{equation}
where $\psi$ is the wave function of the condensate and $g=\hbar c/(\sqrt {2} \rho_0 \xi)$ is proportional to the scattering length (with $c$ the speed of sound, $\rho_0$ the mean mass density, and $\xi$ the healing length); in terms of these variables $m=\hbar/(\sqrt {2} c \xi)$. In dimensionless units, all simulations have $\rho_0=1$, $c=2$, and $\xi$ such that the vortex cores are well resolved by the spatial resolution considered. This results in $\xi=(40\sqrt{2})^{-1}$ in all simulations with $N_x=N_y=256$ spatial grid points, and $\xi=(80\sqrt{2})^{-1}$ in all simulations with $N_x=N_y=512$ grid points so that $\xi k_{max} = 1.5$ where $k_{max}$ is the maximum resolved wavenumber.

The total energy $E_\textrm{tot}$ is a conserved magnitude in the GPE, and can be decomposed into
\begin{equation}
    E_\textrm{tot} = E_k + E_\textrm{int} + E_q,
    \label{eq:E_decomposition}
\end{equation}
where $E_k$ is the kinetic energy, $E_\textrm{int}$ is the internal energy, and $E_q$ is the quantum energy, which are defined respectively as
\begin{equation}
    E_k = \int \frac{(\sqrt{\rho} u)^2}{2} d^3r , \,\,\,\,\,\,\,\,\,\,\,
    E_\textrm{int} = \int \frac{g\rho^2}{2m^2} d^3r , \,\,\,\,\,\,\,\,\,\,\,
    E_q = \int \frac{\hbar^2}{2m^2} (\boldsymbol{\nabla}\sqrt{\rho})^2 d^3r ,
    \label{eq:defEq}
\end{equation}
where $\rho$ is the fluid density and $\boldsymbol{u}$ the fluid velocity, obtained from Madelung's transformation with $\rho = |\psi|^2$ and $\boldsymbol{u} = \hbar \boldsymbol{\nabla}\phi / m$. In this description, quantized vortices correspond to lines with $\rho=0$, with quantum of circulation $\Gamma=h/m$. The Helmholtz decomposition $\sqrt{\rho}\boldsymbol{u} = (\sqrt{\rho}\boldsymbol{u})_{i} + (\sqrt{\rho}\boldsymbol{u})_c$ can be applied to the kinetic energy to further decompose it into incompressible $E_k^i$ and compressible $E_k^c$ kinetic energy components \cite{Nore1997, ClarkdiLeoni2017DualTurbulence}. 
As these energies are quadratic, it is straightforward to construct power spectra from them as
  \begin{equation}
  E_k^{i,c}(k) = \int \frac{1}{2} \left|\widehat{(\sqrt{\rho} u)_{i,c}} \right|^2 k^2 d\Omega_k ,
  \,\,\,\,\,\,\,\,\,\,\,
  E_\textrm{int}(k) = \int \frac{g \left|\widehat{\rho}\right|^2}{2m^2} k^2 d\Omega_k,
  \,\,\,\,\,\,\,\,\,\,\,
  E_q (k) = \int \frac{\hbar^2}{2m^2} \left| \widehat{ (\boldsymbol{\nabla}\sqrt{\rho}) } \right|^2 k^2 d\Omega_k,
  \end{equation}
  where the hat denotes the Fourier transform, and $\Omega_k$ is the solid angle in Fourier space.

The GPE was evolved in time using a fourth-order Runge-Kutta method, and a pseudospectral method to compute spatial derivatives and nonlinear terms \cite{ghost}. Time steps were chosen to satisfy the Courant–Friedrichs–Lewy condition, and resulted in $\Delta t=10^{-3}$ in dimensionles units in simulations with $N_x=N_y=256$ grid points, and in $\Delta t=5\times 10^{-4}$ in simulations with $N_x=N_y=512$ grid points. With these choices, total energy is conserved in all simulations up to the sixth significant digit at $t=10$.

\section{Preparation of the initial conditions} \label{sec:IC&F}

An initial random two-dimensional (2D) flow with a three-dimensional (3D) perturbation is constructed using a Clebsch representation of the incompressible velocity field 
$\boldsymbol{u} = \lambda\boldsymbol{\nabla}\mu-\boldsymbol{\nabla}\phi$,  \citep{Nore1997}, 
where the Clebsch potentials are a superposition of modes
\begin{eqnarray}
\lambda &=& \frac{1}{2k_\lambda} \sum\limits_{k_i=1}^{2k_\lambda} \cos \left\{x \left[k_\lambda \cos \left(\frac{\pi k_i}{2k_\lambda}\right) \right] + y \left[k_\lambda \sin \left(\frac{\pi k_i}{2k_\lambda} \right)\right] + \phi_{k_i} \right\} \times \left\{1 + A_z \cos \left(\frac{2\pi z}{L_z} + \varphi_{k_i}\right) \right\} , \\
\mu &=& \frac{1}{2k_\mu} \sum\limits_{k_j=1}^{2k_\mu} \cos \left\{x \left[k_\mu \cos \left(\frac{\pi k_j}{2k_\mu} \right) \right] + y \left[k_\mu \sin \left(\frac{\pi k_j}{2k_\mu} \right) \right] \right\} ,
\end{eqnarray}
where $\phi_{k_i}$ and $\varphi_{k_i}$ are random phases, the brackets $[\, . \,]$ indicate the integer part of the argument (to satisfy periodicity of each mode), and $ k_z $ is the wavenumber of the perturbation in the $z$ direction. The Clebsch potential $\phi$ is determined by the condition $\boldsymbol{\nabla}\cdot  \boldsymbol{u}=0$. The parameters $ k_\lambda $ and $ k_\mu $ control the initial correlation length of the field, and $A_z$ controls the amplitude of the 3D perturbation. The initial conditions are designed to generate a disordered flow with quantized vortices that have a dominant 2D component, a 3D perturbation (when $A_z \neq 0$), and a correlation length at intermediate scales (or wavenumbers) such as both direct or inverse cascades can develop. As described in \cite{Nore1997, ClarkdiLeoni2017DualTurbulence}, with these potentials an associated initial wavefunction $\psi(x,y,z)$ can be constructed as the product of wavefunctions $\psi_e(\lambda(x,y,z), \mu(x,y,z))$ where the $\psi_e$ have zeros at the zeros of the Clebsch potentials (and thus quantized vortices in the corresponding $x$, $y$, and $z$ coordinates). To reduce the contribution of compressible modes and the initial emission of phonons, before solving the GPE these initial conditions are integrated to convergence using the advective real Ginzburg-Landau equation \cite{Nore1997, ClarkdiLeoni2017DualTurbulence}, which is the imaginary-time propagation of the GPE Galilean transformed to preserve the velocity field $\boldsymbol{u}$. The final result is a wavefunction compatible with the flow $\boldsymbol{u}$ and with minimal sound emission, and which is used as the actual initial condition of the GPE.

\begin{figure}
    \includegraphics[width=.65\columnwidth,trim={20mm 50mm 10mm 60mm},clip]{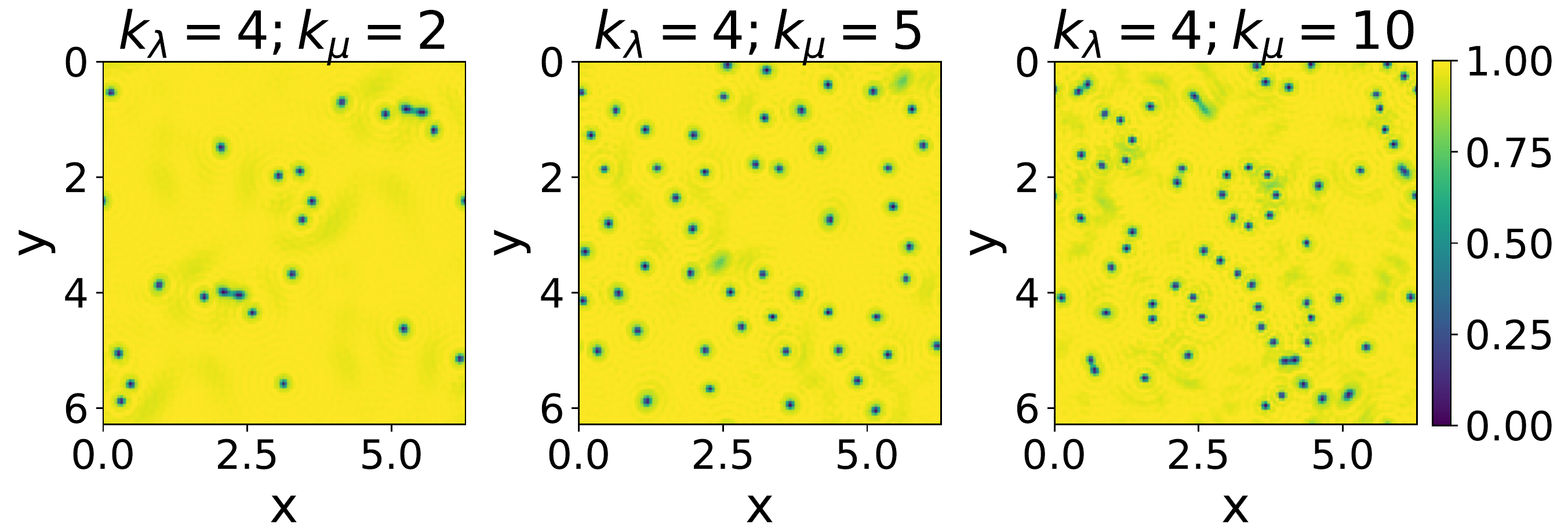}
    \caption{Slices of the mass density in an $xy$ plane for initial conditions with different values of  $k_\lambda$ and $ k_\mu$ and with $A_z=0$. Dark regions correspond to vortex cores. Note the random spatial distribution of vortices, and the change in their number and mean separation as $ k_\mu$ is varied (similar results are obtained when $k_\lambda$ is changed).}
    \label{fig:klam_density}
\end{figure}

\begin{figure}
    \includegraphics[width=.65\linewidth,trim={0 0 0 5mm},clip]{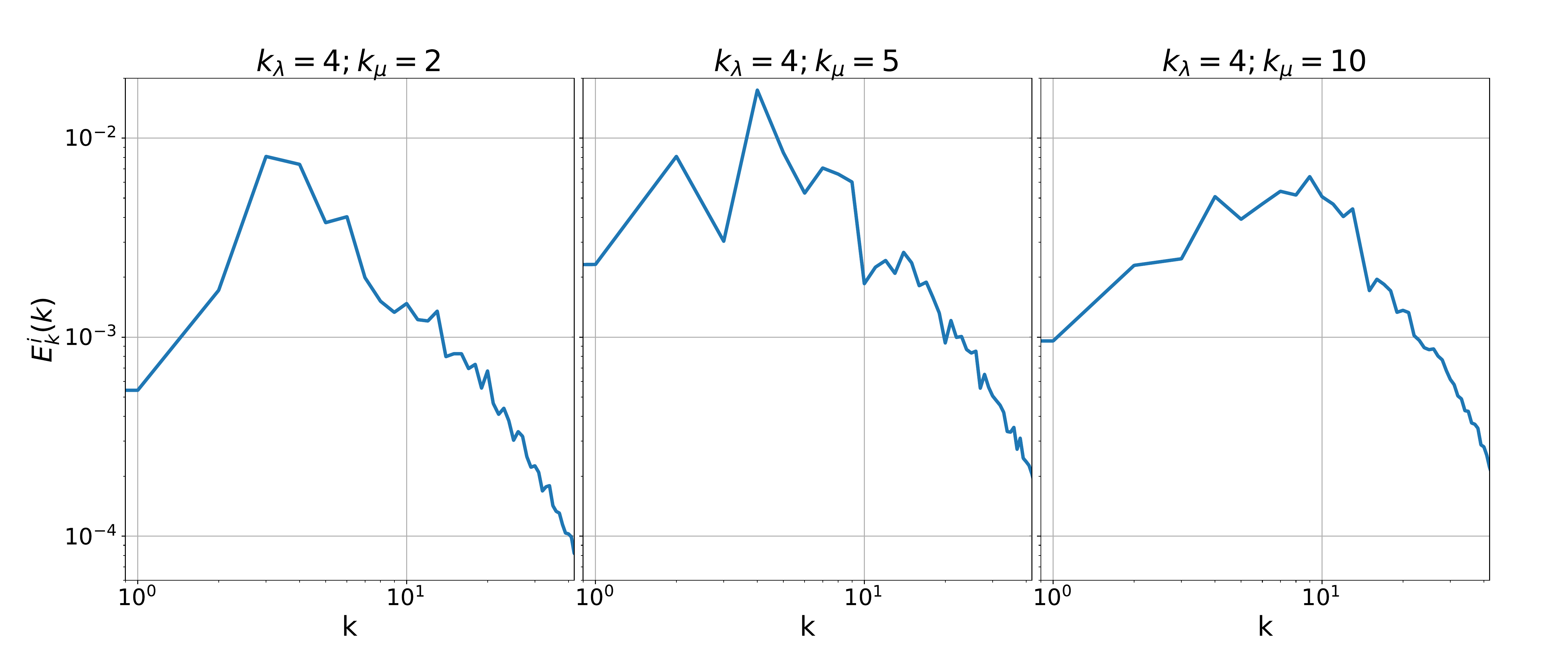}
    \caption{Incompressible kinetic energy spectrum for the initial conditions in Fig.~\ref{fig:klam_density}. The wavenumber corresponding to the maximum of the kinetic energy spectral density changes with $k_\lambda$ and $k_\mu$.}
    \label{fig:klam_spectra}
\end{figure}

As examples of the resulting initial conditions, Fig.~\ref{fig:klam_density} shows slices in an $xy$ plane of the mass density using $128^3$ spatial grid points, for $k_\lambda=4$ and different values of $k_\mu$ (and with $A_z=0$, and thus 2D). Points with zero density are defects that correspond to quantized vortices. As the number of excited modes increases (i.e., as $k_\mu$ increases), more vortices are generated, and the average distance between them decreases. Figure \ref{fig:klam_spectra} shows the spectrum of the incompressible kinetic energy for each of these initial conditions, and it can be seen that the maximum of the spectrum takes place at a wavenumber that increases with $k_\mu$ (i.e., the initial correlation of the flow changes as this quantity is varied) in accordance with the relation $k_{\ell} \sim \sqrt{Mk_0k_{\xi}}$ with $M$ the Mach number and $k_{\xi} \sim 1/\xi$, both of them fixed values, and $k_{\ell} \sim 1/\ell$ the wave number associated with the intervortex distance $\ell$ \cite{Nore1997}. Similar results are obtained when $k_\lambda$ is varied. To consider an initial flow with scale separation such that both direct and inverse energy cascades can develop, all simulations in this study are done with $k_\lambda = 4$ and $k_\mu = 10$, such that the initial energy peaks at $ k_0\approx 10 $.

\begin{figure}
    \centering
    \includegraphics[width=.65\linewidth,trim={20mm 50mm 15mm 60mm},clip]{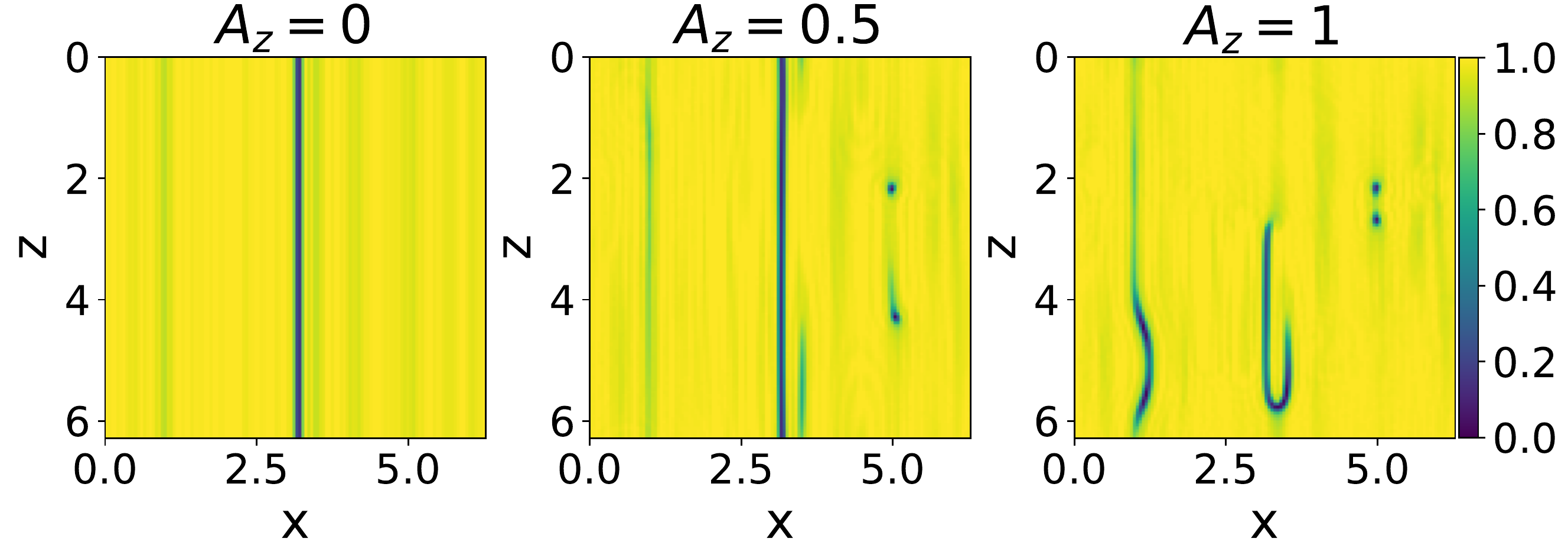}
    \caption{Slices of the mass density in an $xz$ plane for initial conditions with different values of  $A_z$. Dark regions correspond to vortex cores. The configuration is independent of $z$ for $A_z=0$, while the vortex cores become more deformed in the vertical direction as $A_z$ is increased.}
    \label{fig:Az}
\end{figure}

The effect of varying $A_z$ is illustrated in Fig.~\ref{fig:Az}, which shows slices in the $xz$ plane of the mass density for initial conditions with fixed $k_\lambda$ and $k_\mu$, and with different values of $A_z$. For $A_z=0$ vortices are parallel and straight in the $z$ direction, and thus generate a purely 2D flow. As $A_z$ increases the vortices curve until in some cases they can even close on themselves forming rings, generating an initially 3D flow. Videos of these initial conditions in cubic boxes and in thin domains, as well as of their time evolution under the GPE, can be seen as supplemental material.

For the study of the transition in cubic boxes, we considered in simulations with $N_x\times N_y\times N_z=512^3$ grid points values of $A_z=1$, $0.4$, $0.1$, $5\times 10^{-2}$, $4\times 10^{-2}$, $3\times 10^{-2}$, $10^{-2}$, $8\times 10^{-3}$, $6\times 10^{-3}$, $10^{-3}$, $8\times 10^{-4}$, $7\times 10^{-4}$, $6\times 10^{-4}$, $3\times 10^{-4}$, $10^{-5}$, and 0 (even more values of $A_z$, in the same range, were considered in the simulations with $N_x\times N_y\times N_z=256^3$, for a total of 33 simulations at this resolution). In the thin domain case, the 3D perturbation was fixed at $A_z=0.1$, and the aspect ratio was varied to take values $\gamma=1$, $1/2$, $1/4$, $1/8$, $1/10$, $1/12.5$, $1/16$, $1/20$, $1/32$, and $1/64$. As a result, a total of 59 simulations with different parameters was considered for the analysis.

\section{Energy fluxes} \label{sec:EF}

Under the GPE, the dynamics of the system conserves the total energy
\begin{equation}
    \frac{dE}{dt} = 0,
    \label{eq:E_conservation}
\end{equation}
which as a result implies that a detailed balance equation can be written in spectral space as
\begin{equation}
    \frac{dE}{dt}(k) = T(k),
    \label{eq:spectral_conservation}
\end{equation}
where $T(k)$ is the transfer function \cite{rhk_montgo, Mininni2013InverseTurbulence, Alexakis2018CascadesFlows}. In other words, the change of energy at any given wavenumber must correspond to a transfer of this energy to or from this wavenumber to all other wavenumbers. By integrating this equation up to some wavenumber, an energy flux can be defined as
\begin{equation}
    \Pi(k) = -\int\limits_0^k T(k') dk' = -\frac{d}{dt} \int\limits_0^k E(k') dk' = -\frac{dE^<(k)}{dt}.
\end{equation}
Using the decomposition of the energy in Eq.~(\ref{eq:E_decomposition}) and the Helmholtz decomposition, this flux can be further decomposed as
\begin{equation}
    \Pi(k) = \Pi_k^{i}(k) + \Pi_k^{c}(k) + \Pi_\textrm{int}(k) + \Pi_q(k),
    \label{eq:pitot}
\end{equation}
where each component of the flux corresponds to the different energy components. We verified that similar results are obtained when the total energy flux $\Pi(k)$ is used to measure the direction of the cascades, and when the flux of incompressible kinetic energy $\Pi_k^{i}(k)$ is considered instead.

\section{Estimation of vortex Length} \label{sec:VortLength}

In a similar fashion as with the energy, one can define an incompressible momentum power spectrum. The high wavenumber components of this spectrum can be approximated as the sum of the momenta of all the vortices present in the flow, counted individually. This provides an easy way to estimate the total line length of the vortices in the flow. The method is detailed in references \cite{Nore1997, PhysRevA.99.043605}.

\section{Movies} \label{sec:Mov}

The movies provided as supplemental material span the entire time evolution of the flow (from $t=0$ to 10). The 3D renderings of quantized vortices in these movies provide examples of the behavior above and below the critical parameter $A_z^c$ or $\gamma^c$ (or, in other words, 2D-like and 3D-like behavior), and correspond to the following cases:
\begin{itemize}
\item Files side\textunderscore0003.mp4 and side\textunderscore4.mp4 are two examples of vortex evolution in the $3D$ cubic domain (at $512^3$ resolution), respectively with $A_z=0.0003$ ($A_z<A_z^c$) and with $A_z=0.4$ ($A_z>A_z^c$). In the case with $A_z<A_z^c$, note the system remains quasi-2D for a long time, until eventually 3D perturbations grow and dominate the dynamics.
\item Files aniso\textunderscore125.mp4 and aniso\textunderscore03125.mp4 are two examples of vortex evolution in the thin domain, one with $\gamma=0.125$ ($\gamma>\gamma^c$) and the other with $\gamma=0.03125$ ($\gamma<\gamma^c$). In the case with $\gamma<\gamma^c$ the flow remains quasi-2D at all times, showing no vortex reconnection and spatial aggregation of quantized vortices.
\end{itemize}

\end{document}